\documentclass[prl,twocolumn,floatfix,showpacs]{revtex4-1}

\usepackage{graphicx}
\usepackage{amsmath,amssymb}

\newcommand{\avg}[1]{\left\langle #1 \right\rangle}
\newcommand{\eq}[1]{Eq.~(\ref{#1})}
\newcommand{\fig}[1]{Fig.~\ref{#1}}
\newcommand{\bra}[1]{ \left\langle #1 \right |}
\newcommand{\ket}[1]{ \left |#1 \right\rangle}

\newcommand{\abs}[1]{\left| #1 \right|}
\newcommand{\redavg}[1]
	{\left\langle\hspace{-0.7mm}\left\langle #1 \right\rangle\hspace{-0.7mm}\right\rangle}
\newcommand{\comm}[1]{\left[#1\right]}
\def \half{{\frac{1}{2}}}

\def \dperp{{d_\perp}}
\def \sz {\sigma^z}
\def \sp {\sigma^+}
\def \sm {\sigma^-}

\def \nth {\bar n_{\rm th}}
\def \ndr {\bar n_{\rm dr}}

\def \tr{\text{tr}}

\def \dperp{{d_\perp}}

\def \sz {\sigma^z}

\def \sp {\sigma^+}
\def \sm {\sigma^-}
\def \nth {\bar n_{\rm th}}
\def \ndr {\bar n_{\rm dr}}
\def \n {\bar n}
\def \shalf {\tfrac{1}{2}}

\def \Hint {H_{\rm int}}

\begin{document}

\title{Phonon-induced spin-spin interactions in diamond nanostructures:\\
application to spin squeezing}

\date{\today}

\date{\today}
\author{S.~D.~Bennett$^1$}
\author{N.~Y.~Yao$^1$}
\author{J.~Otterbach$^1$}
\author{P.~Zoller$^{2,3}$}
\author{P.~Rabl$^4$}
\author{M.~D.~Lukin$^1$}
\affiliation{$^1$Physics Department, Harvard University, Cambridge, 
Massachusetts 02138, USA} 
\affiliation{$^2$Institute for Quantum Optics and
Quantum Information, Austrian Academy of Sciences, 6020 Innsbruck, Austria}
\affiliation{$^3$Institute for Theoretical Physics,  University of Innsbruck,
6020 Innsbruck, Austria}
\affiliation{$^4$Institute of Atomic and Subatomic Physics, TU Wien,
Stadionallee 2, 1020 Wien, Austria}

\begin{abstract}
We propose and analyze a novel mechanism for long-range
spin-spin interactions in diamond nanostructures.
The interactions between electronic spins,
associated with nitrogen-vacancy centers in diamond, 
are mediated by their
coupling via strain
to the  vibrational mode of a diamond 
mechanical nanoresonator. 
This coupling 
results in
phonon-mediated effective spin-spin interactions that 
can be used to generate squeezed states of
a spin ensemble.
We show that spin dephasing and relaxation can be 
largely suppressed,
allowing for substantial spin squeezing 
under realistic experimental conditions.
Our approach has implications for spin-ensemble
magnetometry, as well as 
phonon-mediated quantum information processing
with spin qubits.   
\end{abstract}

\pacs{ 07.10.Cm, 	
            71.55.-i,     	
            42.50.Dv   	
           }
           
\maketitle

Electronic spins associated with nitrogen-vacancy (NV) centers in 
diamond exhibit long coherence 
times and optical addressability, motivating
extensive 
research on NV-based quantum information  and
sensing applications.
Recent experiments have demonstrated coupling of 
NV electronic spins
to nuclear spins \cite{Jelezko2004,Childress2006},
entanglement with photons \cite{Togan2010},
as well as   single spin \cite{Maze2008,Balasubramanian2008}
and ensemble \cite{Acosta2009,Pham2011}
magnetometry. 
An outstanding challenge is the realization of controlled 
interactions between several NV centers, 
required for quantum gates  
or to generate entangled spin states for quantum-enhanced sensing.
One approach toward this goal  is to
couple  NV centers
to a resonant  optical \cite{Englund2010,Faraon2011}
or mechanical \cite{Rabl2010,Arcizet2011,Kolkowitz2012} mode;
this  is particularly appealing in light of
rapid progress 
in the fabrication of diamond nanostructures with 
improved optical and mechanical properties 
\cite{Zalalutdinov2011,Burek2012,Hausmann2012,Ovartchaiyapong2012,Tao2012}.

\begin{figure}[b]
\centering
	\includegraphics[width=0.45\textwidth]{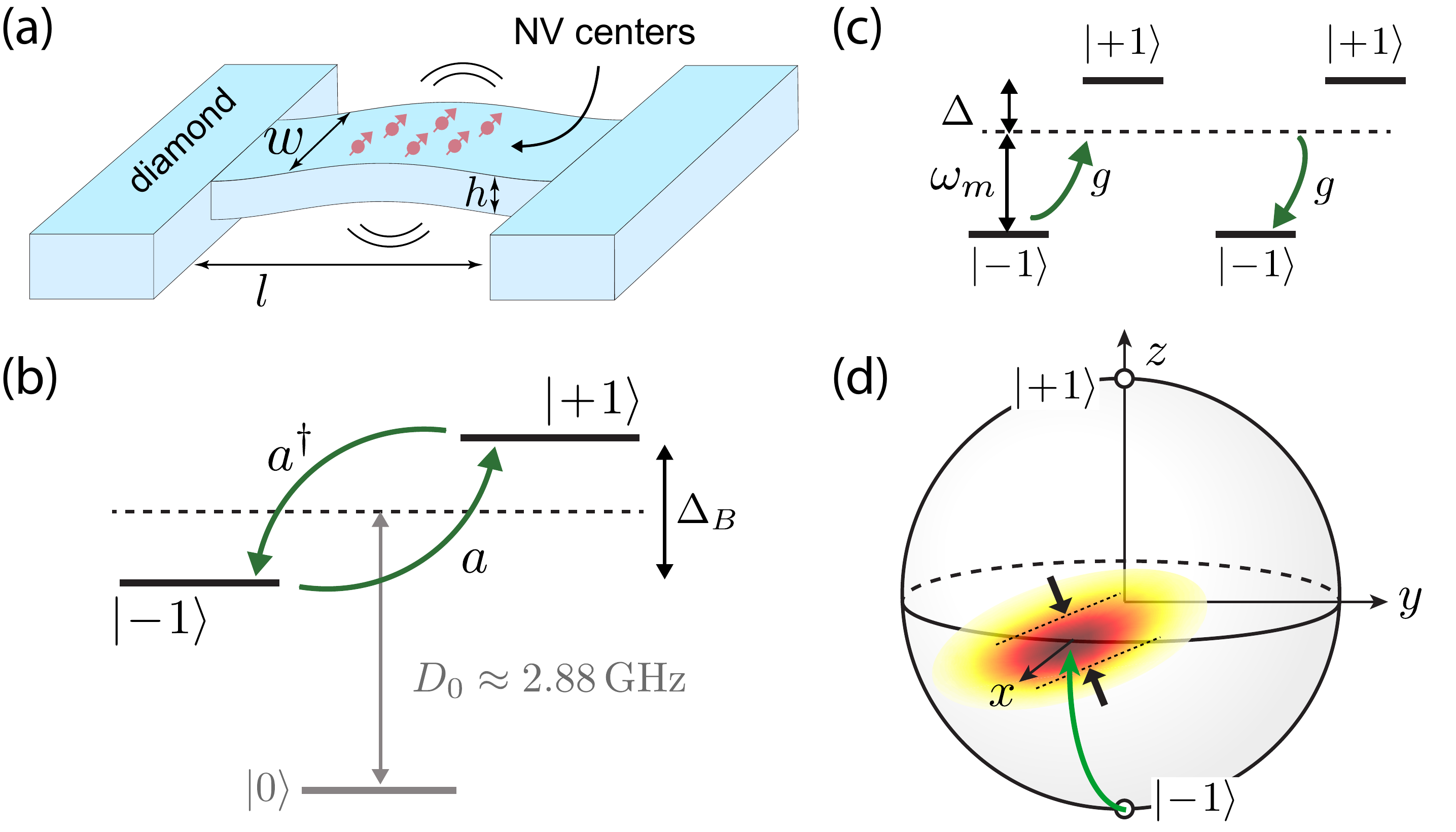}
\caption{
(a) All-diamond doubly clamped mechanical resonator
with an ensemble of embedded NV centers.
(b) Spin triplet states of the NV electronic ground state.
Local perpendicular strain induced by beam bending mixes 
the $\ket{\pm 1}$ states.
(c) A collection spins in the two-level
subspace $\{ \ket{+1},\ket{-1} \}$ is off-resonantly coupled to a common mechanical mode giving rise to effective spin-spin interactions.
(d) Squeezing of the spin uncertainty distribution
of an NV
ensemble.}
\label{fig:cartoon}
\end{figure}

In this Letter, we describe a new approach for
effective spin-spin interactions between NV centers
based on  strain-induced coupling 
to a  vibrational mode of a diamond 
resonator.
We consider
an ensemble of NV centers embedded 
in a single crystal diamond nanobeam,
as depicted in \fig{fig:cartoon}a.
When the beam flexes, it  strains 
the diamond lattice
which in turn couples directly to the
spin triplet states in the NV
electronic ground state \cite{Maze2011,Doherty2012}.
For a thin beam of length $L \sim 1\ \mu$m, 
this strain-induced spin-phonon coupling 
can 
allow for coherent effective
 spin-spin interactions
mediated by
virtual phonons.
Based on these effective interactions,
we explore the possibility to generate 
spin squeezing of an NV ensemble
embedded in the nanobeam.
We account for spin dephasing and mechanical dissipation, 
and describe how spin echo techniques 
and mechanical driving can be used to suppress
the dominant decoherence processes while 
preserving the coherent spin-spin interactions.  
Using these techniques we find that significant 
spin squeezing can be achieved with realistic experimental parameters. 
Our results have implications for NV ensemble magnetometry,
and provide a new route toward controlled long-range
spin-spin interactions.

{\it Model.---}The
 electronic ground state of the 
negatively charged NV center 
is a spin $S=1$ triplet with spin states labeled by 
$\ket{m_s= 0,\pm 1}$ as 
shown in \fig{fig:cartoon}b.
In the presence of external electric and magnetic fields 
$\vec E$ and $\vec B$, the Hamiltonian for a single NV is  
$(\hbar=1)$ \cite{Doherty2012}
\begin{equation}
\label{eq:Hnv}
\begin{split} 
	H_{\rm NV}= &( D_0 + d_\parallel E_z ) S_z^2  + 
	\mu_B g_s   \vec S \cdot \vec B  \\
	&- d_\perp \left[E_x( S_xS_y+S_y S_x) + E_y (S_x^2-S_y^2)\right],
\end{split}Ê
\end{equation} 
where $D_0 / 2\pi \simeq 2.88$ GHz is the zero field splitting, 
$g_s \simeq 2$, $\mu_B$ is the Bohr magneton,
and $d_\parallel$  ($d_\perp$) is
the ground state electric dipole moment in the direction
parallel (perpendicular) to the NV axis 
\cite{Vanoort1990,Dolde2011}.

Motion of the diamond nanoresonator changes 
the local strain at the position of the NV center, which results 
in an effective, strain-induced electric field  \cite{Doherty2012}.
We are interested in the near-resonant coupling of 
a single resonant mode  of the  nanobeam to
the $\ket{\pm 1}$ transition of the
NV, with Zeeman splitting
$\Delta_B=  g_s \mu_B B_z/\hbar$,
as shown in \fig{fig:cartoon}b,c.
The perpendicular component
of strain $E_\perp$
mixes the $\ket{\pm 1}$ states.
For small beam displacements, the
strain is linear in its  position
and we write 
$E_\perp = E_0 (a + a^\dagger)$,
where $a$ is the destruction operator of 
the resonant mechanical mode 
of frequency $\omega_m$,
and
$E_0$ is the perpendicular strain 
resulting from the zero point motion
of the beam.
We note that the parallel 
component of strain  
shifts both states $\ket{\pm 1}$  relative 
to  $\ket{0}$ \cite{Acosta2010};
however, with near-resonant coupling
$\Delta = \Delta_B - \omega_m \ll D_0$
and preparation in 
the $\ket{\pm 1}$ subspace,
the state $\ket{0}$ remains unpopulated
and parallel 
strain plays no role in what follows.
Within this two-level subspace, the interaction 
of each NV is 
$H_i = g \left( \sigma^+_i a + a^\dagger \sigma^-_i \right)$,
where
$\sigma^\pm_i = \ket{\pm 1}_i\bra{\mp 1}$
is the Pauli operator of the
$i$th NV center and
$g$ is the single phonon coupling strength.
For many NV centers 
we introduce  collective spin operators,
$J_z = \frac{1}{2} \sum_i \ket{1}_i\bra{1} - \ket{-1}_i\bra{-1}$ and 
$J_{\pm} = J_x \pm i J_y = \sum_i \sigma^\pm_i$,
which satisfy the usual angular momentum
commutation relations. 
The total system Hamiltonian can then be written as 
\begin{equation}
\label{eq:Hstart}
	  H = \omega_m a^\dag a + \Delta_B J_z + g \left( a^\dagger J_-  + a J_+ \right),
\end{equation}
which describes a Tavis-Cummings type interaction 
between an ensemble of spins and a single mechanical mode.
In \eq{eq:Hstart} we have assumed 
uniform coupling of each
spin to the mechanical mode for simplicity.
In general the coupling may be nonuniform and
we 
discuss this  further below.

To estimate the coupling strength $g$, we 
calculate the strain for a given
mechanical mode  
and use the
experimentally obtained stress coupling of 
$0.03$ Hz Pa$^{-1}$  in the NV ground state 
\cite{Togan2011,si}.
We take
a doubly clamped diamond beam (see \fig{fig:cartoon}a)
with dimensions $L \gg w,h$ such that Euler-Bernoulli 
thin beam elasticity theory
is valid \cite{LL}. 
For NV centers located near the surface of the beam we obtain
\cite{si}
\begin{equation}
\label{eq:GSstress}
	\frac{g}{2\pi} \approx  180
	 \left( \frac{\hbar}{L^3 w  \sqrt{\rho E}} \right)^{1/2}
	 {\rm GHz},
\end{equation}
where $\rho$ is the mass density and 
$E$ is the Young's modulus of diamond.
For a beam of dimensions $(L,w,h)=(1,0.1,0.1)\,\mu$m 
we obtain a vibrational frequency $\omega_m/2\pi \sim 1$ GHz 
and  coupling  $g/2\pi \sim 1$ kHz. 
While this is smaller than  
the strain coupling 
$g_{e}/2\pi\approx 10$ MHz
expected for electronic excited states of defect 
centers \cite{Habraken2012,Soykal2011}
or quantum dots \cite{Wilson-Rae2004},
we benefit  from the much longer 
spin coherence time $T_2$
in the ground state.
An important figure of merit is
the single spin cooperativity 
$\eta = \frac{g^2 T_2}{ \gamma \nth}$,
where $\gamma=\omega_m/Q$ is the mechanical damping rate 
and $\bar n_{\rm th}  =  (e^{\hbar \omega_m/k_BT} - 1)^{-1}$
is the equilibrium phonon occupation number at temperature $T$.  
Assuming $Q = 10^6$, $T_2 = 10$ ms and $T = 4$ K,  we obtain a
single spin cooperativity of $\eta \sim 0.8$.
This can be further increased by  reducing the dimensions
of the nanobeam and
operating at  lower temperatures.


{\it Spin squeezing.---}In
the dispersive
regime, $g \ll \Delta = \Delta_B - \omega_m$, 
virtual excitations of the mechanical mode result in 
effective interactions between the otherwise decoupled spins. 
In this  limit,
$H$
can be approximately
 diagonalized 
by the
transformation $e^R H e^{-R}$ with
$R = \frac{g}{\Delta}  \left( a^\dagger J_- - a J_+ \right)$.
To  order  $(g/\Delta)^2$  this 
 yields an effective Hamiltonian,
\begin{equation}
\label{eq:Heff}
	H_{\rm eff} =\omega_m a^\dag a +   \left( \Delta_B +\lambda a^\dag a \right)  J_z
	+ \frac{\lambda}{2} J_+J_-,	
\end{equation}
where  $\lambda = 2 g^2/\Delta$ is the
phonon-mediated  spin-spin coupling  strength.
Rewriting 
$J_+ J_- =  {\bf{J}} ^2 - J_z^2 + J_z$, 
and provided the total angular momentum $J$ 
is conserved,
we obtain a  term $\propto J_z^2$  corresponding
to the one-axis twisting  Hamiltonian \cite{Kitagawa1993}.

To 
generate a spin squeezed state, we  initialize the ensemble 
in a coherent spin state (CSS) $\ket{\psi_0}$ along the
$x$ axis of the collective  Bloch sphere.
The CSS satisfies $J_x \ket{\psi_0} = J \ket{\psi_0}$ 
and has equal 
transverse variances, $\avg{J_y^2} = \avg{J_z^2} = J/2$.
This can be achieved using  optical 
pumping and global rotations of the spins with microwave fields
\cite{Taylor2008}.  
The squeezing term $\propto J_z^2$ describes a 
precession of the collective spin about the $z$ axis at
a rate proportional to $J_z$, resulting in a shearing of
the uncertainty distribution
and a reduced spin variance in one direction
as shown in \fig{fig:cartoon}d. 
This is quantified by
the squeezing parameter \cite{Wineland1992,Ma2011},
\begin{equation}
\label{eq:xi}
	\xi^2 = \frac{2J \avg{\Delta J_{\rm min}^2}}{\avg{J_x}^2},
\end{equation}
where $\avg{\Delta J_{\rm min}^2} 
	= \frac{1}{2}
	\left( V_+ - \sqrt{ V_-^2 + V_{yz}^2}
	\right)$ is
the minimum spin uncertainty with
$V_\pm = \avg{J_y^2 \pm J_z^2}$
and $V_{yz} = \avg{J_y J_z + J_z J_y}/2$.
The preparation of a spin squeezed state,
characterized by  $\xi^2<1$,
has direct implications for NV 
ensemble magnetometry 
applications, since it would enable magnetic field 
sensing with a 
precision below the projection noise limit
\cite{Wineland1992}.

We now consider spin squeezing
in the presence of realistic  decoherence. 
In addition to the coherent dynamics described
by $H_{\rm eff}$, 
we account for mechanical dissipation and
spin dephasing using  
a master equation \cite{si}
\begin{align}
\label{eq:ME}
	\dot \rho = & - i \left[ - \frac{\lambda}{2} J_z^2 + \left( \Delta_B + 
	\lambda a^\dagger a \right) J_z ,\rho \right]
	+ \frac{1}{2T_2} 
	\sum_i \mathcal{D}	[\sigma_z^i]\rho \nonumber \\
	&+ {\Gamma_\gamma} (\nth +1)\mathcal{D}[J_-] 
	+ {\Gamma_\gamma} \nth \mathcal{D}[J_+],
\end{align} 
where $\mathcal{D}[c] \rho= c\rho c^\dag - \frac{1}{2} \left( c^\dag c \rho + \rho c^\dag c\right)$
and the single spin dephasing
$T_2^{-1}$ is 
assumed to be
Markovian for simplicity (see below).
Note that we 
absorbed a shift of $\lambda/2$ into $\Delta_B$, and
ignored
single spin relaxation as $T_1$ can 
be several minutes at low temperatures
\cite{Jarmola2012}.
The second line 
describes collective 
spin relaxation induced by mechanical dissipation,
with
$\Gamma_\gamma=\gamma g^2/\Delta^2$.
Finally, the phonon number $n = a^\dagger a$
shifts the spin frequency,  acting as an effective fluctuating 
magnetic field which leads to additional dephasing.

\begin{figure}[tb]
	\includegraphics[width=0.45\textwidth]{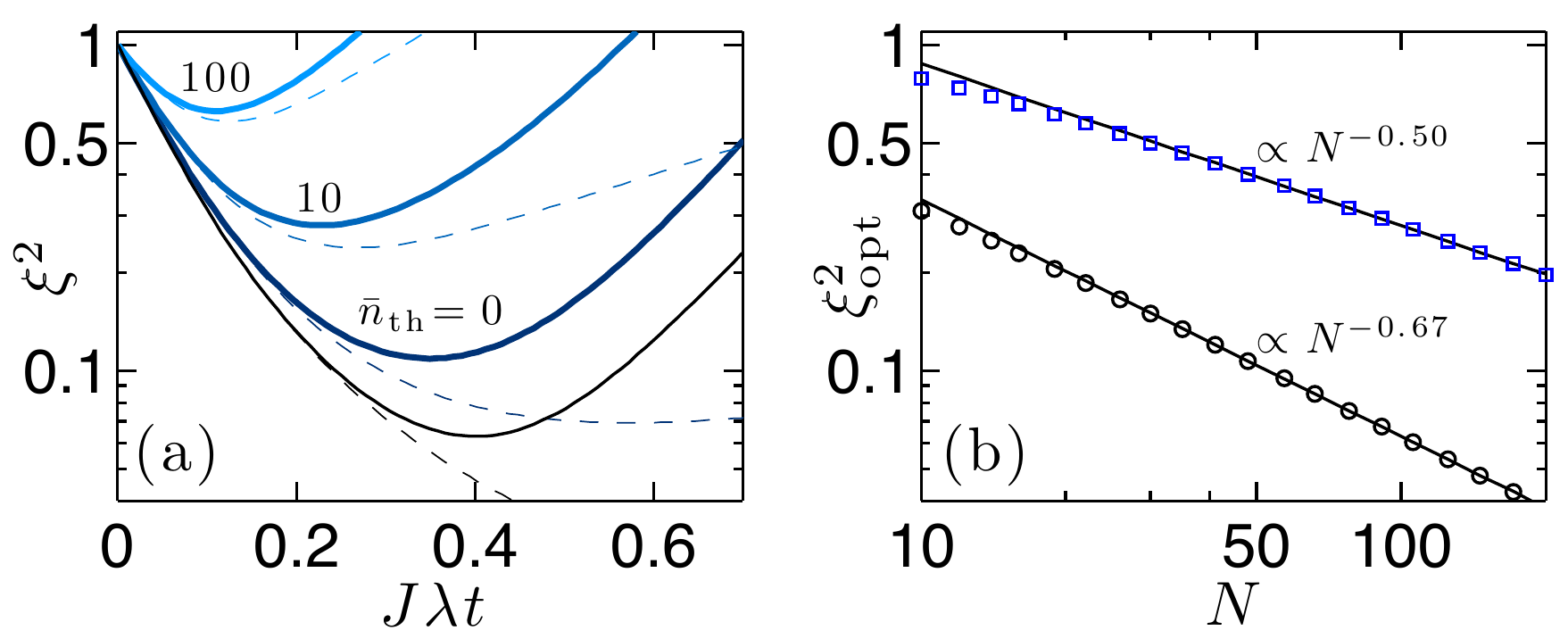}	
\caption{(a) Spin squeezing parameter 
versus scaled precession time with $N = 100$ spins.
Solid blue lines show the calculated squeezing parameter
for   $T_2 = 10$ ms
and values of $\nth$ as shown.
For each curve, we optimized the detuning $\Delta$
to obtain the optimal squeezing.
Blue dashed lines are calculated
from the linearized equations
for the spin operator averages.
Black solid (dashed) line shows exact (linearized)
unitary squeezing.
(b) Optimal squeezing versus number of spins.
Lower (upper) red line shows power law fit
for $\nth = 1$ (10)
and $T_2 = 1$ (0.01) s.
The detuning $\Delta$ is optimized for each point.
Other parameters in both plots are
$\omega_m/2\pi = 1$ GHz,
$g / 2\pi = 1$ kHz,
$Q = 10^6$.
}
\label{fig:nodrive}
\end{figure}

Let us for the moment 
ignore
fluctuations of the phonon number $n$;
we address these in detail below.
Starting from the CSS $|\psi_0\rangle$, we 
plot the squeezing parameter 
in \fig{fig:nodrive}a 
for an ensemble of $N = 100$ spins
and several values of $\nth$,
in the presence of  dephasing $T_2^{-1}$
and collective relaxation $\Gamma_\gamma$.
Here we calculated $\xi^2$
by solving \eq{eq:ME} using an approximate numerical approach
 treating $\Gamma_\gamma$ and $T_2$  separately,
 and verified that the approximation agrees  with exact results
 for small $N$ \cite{si}.
To estimate  the
minimum squeezing, 
we  linearize the equations of motion
for the averages and variances of
the collective spin operators (see dashed lines in \fig{fig:nodrive}a).
From these linearized equations, 
in the limits of interest, $J \gg 1$,
 $\nth \gg 1$
and to leading order in both
sources of decoherence, we 
obtain approximately
\begin{equation}
\label{eq:xiSqApprox}
	\xi^2 \simeq \frac{4  \Gamma_\gamma \nth}{J \lambda^2  t }
	+ \frac{t}{T_2}.
\end{equation} 
Optimizing $t$ and the detuning $\Delta$,
we obtain the 
optimal squeezing parameter,
\begin{equation}
\label{eq:xiOpt}
	\xi^2_{\rm opt} \simeq \frac{2}{\sqrt{ J \eta}},
\end{equation}
at time $t_{\rm opt} = T_2 / \sqrt{J \eta}$,  
similar to results 
for  atomic systems \cite{SchleierSmith2010,Leroux2010,Leroux2012}. 
Note that for non-Markovian dephasing,
the scaling is
even more favorable \cite{Marcos2010}.
In \fig{fig:nodrive}b we plot the
scaling of the squeezing parameter with $J$
for small but finite decoherence,
and find agreement with \eq{eq:xiOpt}.
For  comparison we also plot
the unitary result
in the absence of decoherence,
scaling as $\xi^2_{\rm opt} \sim J^{-2/3}$ and 
limited by the Bloch sphere curvature 
\cite{Kitagawa1993}.

{\it Phonon number fluctuations}.---In 
\eq{eq:Heff} we see that the phonon number
$n = a^\dagger a$ couples to $J_z$,
leading to additional dephasing due to
thermal number fluctuations.
On the other hand, this same coupling can also
lead to additional
spin squeezing from cavity feedback,
by driving the mechanical mode 
\cite{SchleierSmith2010,Leroux2010,Leroux2012}. 
In the following, we 
consider a twofold approach to mitigate
thermal spin dephasing while preserving the optimal
squeezing.
First, we apply a sequence of global spin echo control
pulses to suppress dephasing
from low-frequency thermal fluctuations.
This also
extends the effective coherence time
$T_2$ of single NV spins
\cite{Taylor2008}.
Second, we consider driving the
mechanical mode, and identify conditions
when this results in a net improvement
of the squeezing.

\begin{figure}[tb]
	\includegraphics[width=0.45\textwidth]{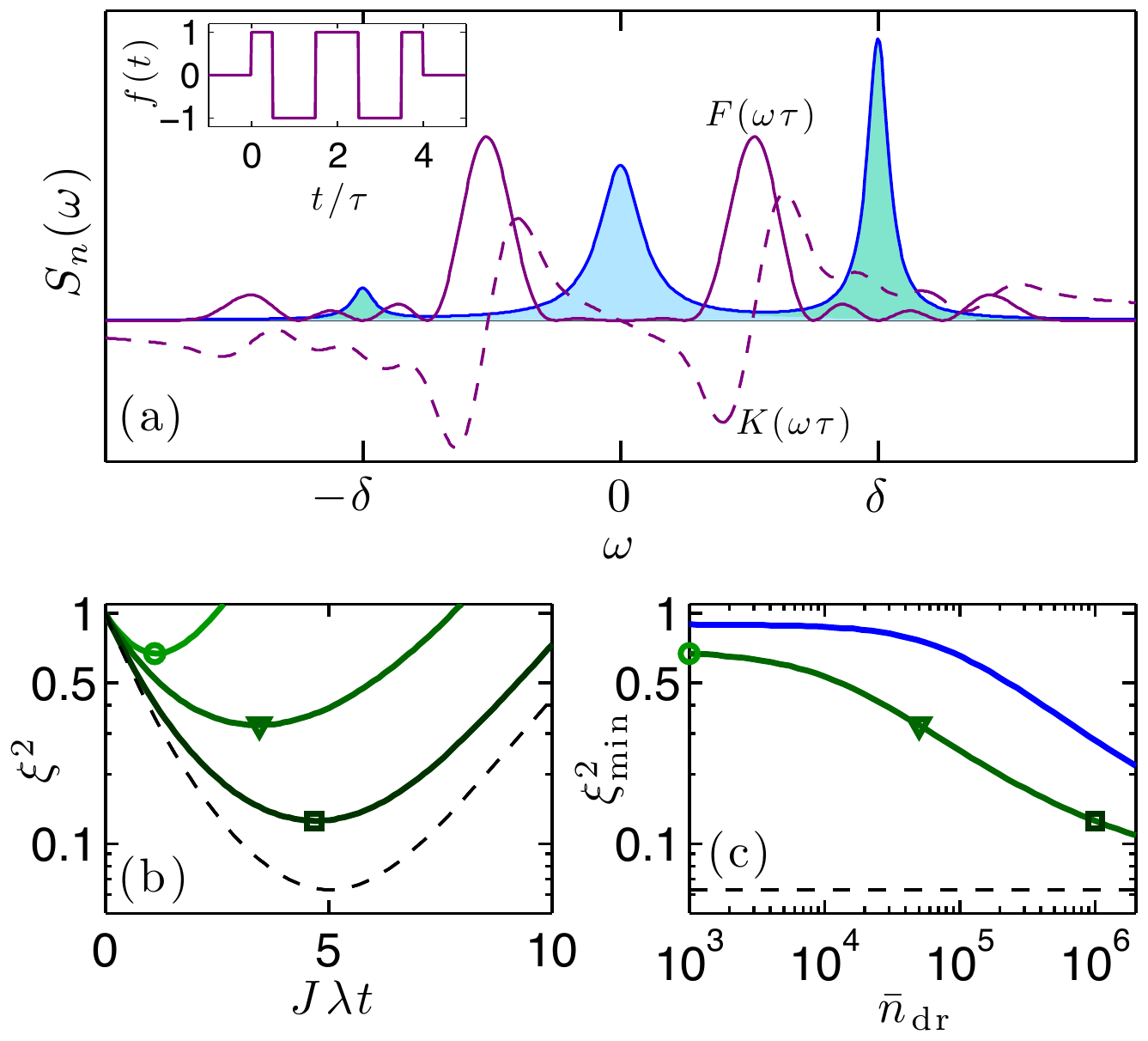}	
\caption{
(a) Number fluctuation spectrum of thermal 
driven oscillator.
Center (blue) peak is purely thermal while
side (green) peaks are due to detuned drive.
Solid (dashed) purple line shows filter function $F$ ($K$) for
$M = 4$ pulses.
Inset: corresponding function $f(t)$ for $M = 4$.
(b) Solid green curves show squeezing 
parameter versus precession time
for $\nth = 10$ and
$\ndr = 10^3, 5 \times 10^4, 10^6$ (top to bottom).
Dashed black line shows unitary squeezing.
 (c) Minimum squeezing versus 
drive strength
for $\nth = 50,10$ (top to bottom).
Symbols mark corresponding points with (b).
Dashed black line shows unitary squeezing.
Parameters in (b) and (c) are
$M = 4$,
$g/2\pi = 1$ kHz, $T_2 = 10$ ms, $N = 100$,
$\omega_m/2\pi = 1$ GHz, $Q = 10^6$.
}
\label{fig:drive}
\end{figure}

To simultaneously account for thermal
dephasing, driven feedback squeezing,  and
 spin control pulse sequences,
we write the interaction term in \eq{eq:Heff}
in the so-called
``toggling frame" \cite{Haeberlen1968},
\begin{equation}
\label{eq:Hint}
	H_{\rm int}(t) = \lambda J_z  f(t) \delta n(t).
\end{equation}
The function $f(t)$ periodically inverts
the sign of the interaction as shown in
the inset of \fig{fig:drive}a, 
describing the inversion of 
the collective spin
$J_z \rightarrow -J_z$ with each $\pi$ pulse of the spin echo sequence.
Phonon number fluctuations are described by
$\delta n(t)=n(t)-\bar n$, where  $\bar n$
is the mean phonon number and
 we have
omitted an average frequency shift proportional to $\bar n$
in \eq{eq:Hint}. 
The number   fluctuation spectrum 
$S_n(\omega) = \int dt e^{i\omega t} \avg{\delta n(t) \delta n(0)} $ is
plotted in \fig{fig:drive}a
for a  driven oscillator coupled to a thermal 
bath \cite{si}.

We  calculate the required spin moments within the 
Gaussian approximation for  phonon number fluctuations,
and obtain \cite{si}
\begin{equation}
\label{eq:Jp}
	\avg{J_+(t)} = e^{-\chi} \avg{e^{-i \mu (J_z - 1/2)} J_+(0)},
\end{equation}
and similar results for 
$\avg{J_+^2(t)}$ and $\avg{J_+(t)J_z(t)}$.
In \eq{eq:Jp} the dephasing parameter $\chi$ and
effective squeezing via
 $\mu$ are given by
\begin{align}
	\chi &=
	\lambda^2 \int \frac{d\omega}{2\pi}
	\frac{F(\omega \tau)}{\omega^2}
	\bar S_n(\omega),
	\label{eq:chi}
	\\
	\mu &=
	\lambda^2 \int \frac{d\omega}{2\pi}
	\frac{K(\omega \tau)}{\omega^2}
	 A_n(\omega),
	 \label{eq:mu}
\end{align}
where $\bar S_n(\omega) = \left( S_n(\omega) + S_n(-\omega) \right) / 2$ 
and
$A_n(\omega) = \left( S_n(\omega) - S_n(-\omega) \right)/ 2$. 
The filter function
$F(\omega\tau) = \frac{\omega^2}{2} \left| \int dt e^{i\omega t} f(t) \right|^2$
describes the effect of the spin echo pulse sequence 
with time $\tau$ between $\pi$ pulses
\cite{Martinis2003,Uhrig2007,Cywinski2008}. 
The function $K(\omega\tau)$ plays the analogous role
for the
effective squeezing described by $\mu$, and is
related to $F$ by a Kramers-Kronig relation \cite{si}. 
We plot $K$ and $F$ for a sequence of $M=4$ 
pulses  in \fig{fig:drive}a.

\emph{Discussion.---}We now consider the impact of 
thermal fluctuations  on the achievable  squeezing.
The  noise spectrum 
$S_n(\omega)=2\gamma \nth(\nth+1)/(\omega^2+\gamma^2)$ is symmetric 
around $\omega=0$.
Without spin echo control pulses, 
this low frequency noise results in  nonexponential 
decay of the spin coherence, 
$\chi_0(t) = \frac{1}{2} \lambda^2 \nth^2 t^2$ (with $\nth \gg 1$),
familiar from 
 single qubit decoherence
\cite{Taylor2008,deSousa2009}. 
The inhomogeneous thermal dephasing time
is  $T_2^* \simeq \sqrt{2} / \lambda \nth$,  
severely limiting the possibility of spin squeezing. 
In particular, at time $t = t_{\rm opt}$ 
we find that
 squeezing is prohibited when 
$\nth > \sqrt{J}$ \cite{si}.
However, one can overcome this low frequency thermal noise using spin echo. 
By  applying a sequence of $M$ equally spaced 
global $\pi$-pulses to the spins during 
 precession of total time $t$, we obtain 
$\chi_{\rm th} \sim \lambda^2 \gamma \nth^2 t^3 / M^2$, suggesting that
thermal dephasing can be made negligible relative to both  $\Gamma_\gamma$
and $T_2^{-1}$.
For a sufficiently large number of pulses,
$M \gg \nth \sqrt{ \gamma T_2}$, we recover 
the optimal squeezing in Eqs.~(\ref{eq:xiSqApprox}) and (\ref{eq:xiOpt}).


Adding a mechanical drive can 
further enhance squeezing via feedback;
however,   it  also increases phonon number fluctuations,
contributing to additional dephasing. 
We  consider  a detuned external drive 
of frequency
$\omega_{\rm dr}=\omega_m+\delta$, leading to
two additional 
peaks 
in $S_n(\omega)$
at  $\omega=\pm \delta$,
as shown in \fig{fig:drive}a.
The area
under the left [right] peak scales as 
$\ndr\nth$ [$\ndr(\nth+1)$],
where $\ndr$ is the 
mean phonon number due to the 
drive at zero temperature.
The symmetric and antisymmetric
parts of this noise contribute
to dephasing and squeezing as described
by Eqs.~(\ref{eq:chi}) and (\ref{eq:mu}).
Choosing the interval $t/M = 2\pi / \delta$ between $\pi$ pulses,
we obtain additional dephasing
$\chi_{\rm dr} \simeq 
\left( \frac{\lambda}{\delta} \right)^2  \ndr \nth \gamma t$
and effective squeezing with
$\mu \simeq \frac{\lambda^2}{\delta} \ndr t$. 
In the limit  $\ndr \gg \nth$,
the effects of the drive
dominate
over $\chi_{\rm th}$ and $\Gamma_\gamma$
and we recover the ideal scaling given in  
\eq{eq:xiOpt},
even with a small number of echo pulses. 
This is shown in  \fig{fig:drive}b,c
where we see  that
the optimal squeezing improves
with increasing $\ndr$
for a fixed number of pulses  $M = 4$.

Finally, we discuss our assumption of
uniform 
coupling strength $g$ in \eq{eq:Hstart}.
This is an important practical
issue, as
we expect the coupling to individual spins
to be inhomogeneous in experiment
due to the spatial variation
of strain in the beam.
Nonetheless, 
even with  nonuniform coupling,
we still obtain squeezing of a
collective spin
with a reduced effective
total spin $J_{\rm eff} < J$,
provided $J \gg 1$. 
First, we note that inhomogeneous magnetic fields
resulting in nonuniform detuning are compensated
by spin echo.
Second, for a distribution of coupling strengths $g_i$, the 
effective length of the collective spin 
is $ \sum_i g_i /\sqrt{ \sum_i g_i^2}$ for 
the direct squeezing term, and 
$\sum_i g_i^2 /\sqrt{ \sum_i g_i^4}$ 
for feedback squeezing with a mechanical drive.   
Similar conclusions were reached
in atomic and nuclear systems
\cite{SchleierSmith2010,Leroux2010,Leroux2012,Rudner2011}. 
In the case of direct squeezing, it is important that
the sign of the $g_i$'s is the same to avoid cancellation;
this is automatically achieved by using NV
centers implanted on the top of the beam.
For  beam dimensions $(1,0.1,0.1)\,\mu$m 
analyzed above, 
we estimate that $N \sim 200$ NV centers can be 
embedded without being perturbed by 
direct magnetic dipole-dipole interactions. 
A reduction of the effective spin length by factor  $\sim 2$ still 
leaves $N_{\rm eff} \sim 100$,
sufficient to observe spin squeezing.

{\it Conclusions.---}We have shown that
direct spin-phonon coupling in diamond
can be used to prepare
spin squeezed  states of an NV ensemble
embedded in a nanoresonator,
even in the presence of dephasing and
mechanical dissipation.
With  further reductions
in temperature,  beam dimensions,
and  spin decoherence rates,
the regime of large single spin cooperativity
$\eta \gg 1$ could be achieved. 
This would  allow for coherent phonon-mediated
interactions and quantum gates between two spins
embedded in the same resonator
via $H_{\rm int} = \lambda \left( \sp_1 \sm_2 +  {\rm h.c.} \right)$,
and coupling over larger distances 
by phononic channels \cite{Habraken2012}.

\emph{Acknowledgments.---}The 
authors gratefully acknowledge 
discussions with 
Shimon Kolkowitz and
Quirin Unterreithmeier.    
This work was supported by NSF, CUA, DARPA, NSERC, HQOC, DOE,
the Packard Foundation, the EU project AQUTE and the Austrian Science Fund (FWF) through SFB FOQUS and the START grant Y 591-N16.

\newpage
\section{Supplemental information}

\subsection{Coupling strength}

We assume that the NV axis is aligned
with both the magnetic field and with the
direction of beam deflection, so that the longitudinal
strain due to deflection
is entirely perpendicular to the NV axis.
From experiment \cite{Togan2010supp},
the splitting of the $\ket{\pm 1}$ states with stress
is $\sim 0.03$ Hz/Pa.
We convert this into 
the deformation potential coupling
frequency,
$\Xi / 2\pi\hbar = 36$ GHz,
using the Young's modulus of diamond, $E = 1200$ GPa.
Next we calculate the strain at the NV center
using elasticity theory.
The equation for the bending mode of a thin beam   is
\begin{equation}
	\rho A \partial_t^2 \phi(z,t) = - E I \partial_z^4 \phi(z,t)
\end{equation}
where $\phi$ is the transverse
displacement in the $x$ direction
and $z$ is along the beam.
Here $\rho$ is the mass density,
$A$ is the cross sectional area, and
$I = w t^3 / 12$ is the moment of inertia.
The solutions are of the form
$\phi(z,t) = u(z)  e^{ - i \omega t}$ where
\begin{equation}
\begin{split}
	u(z) = \frac{1}{\sqrt{N}} \Big[&
	\cos kz - \cosh kz
	\\ &- \frac{(\cos kL - \cosh kL)}{(\sin kL - \sinh kL)} (\sin kz - \sinh kz )
	\Big],	
\end{split}
\end{equation}
which satisfies the
boundary conditions 
$u(0) = u'(0) = u(L) = u'(L) = 0$ for a doubly
clamped beam.
The allowed wavenumbers $k_n$ 
are given
by the solutions of $\cos kL \cosh kL = 1$,
and
the corresponding eigenfrequencies are
\begin{equation}
\label{eq:freq}
	\omega_n = k_n^2 \sqrt{\frac{EI}{\rho A}}.
\end{equation}
We normalize the  modefunction 
of the fundamental mode $u_0(z)$ 
by setting the free energy stored
in the beam to  the zero point energy,
\begin{equation}
	W = \frac{1}{2} EI \int_0^L dz
	\left( \frac{\partial^2 u_0}{\partial z^2} \right)^2
	= \frac{\hbar \omega_0}{2}.
\end{equation}
Integrating by parts we obtain the normalization condition,
\begin{equation}
	\int_0^L dz u_0^2 = \frac{\hbar}{\rho A \omega_0}.
\end{equation}
If the NV center lies at the midpoint along the beam, $z = L/2$, 
and at a distance
$r_0$ from the neutral axis of the beam, the strain due to the
zero point motion of the fundamental mode is
\begin{equation}
	\epsilon_0 =  - r_0 \partial_z^2 u_0(L/2) 
	\sim r_0 \sqrt{\frac{\hbar}{\rho A \omega_0}} \frac{27}{L^{5/2}}
	\sim 5 \frac{2r_0}{t} \sqrt{\frac{\hbar }{L^3 w  \sqrt{E \rho} }} ,
\end{equation}
where we \eq{eq:freq} and $k_0 \sim 4.73 / L$ for the fundamental mode.
The coupling strength is given by the deformation potential and
the strain due to zero point motion,
\begin{align}
	\frac{ g}{2\pi} =  \frac{ \Xi}{2\pi \hbar} \epsilon_0  \sim  180\ {\rm GHz} \cdot
	\frac{2r_0}{t}
	 \sqrt{\frac{\hbar}{L^3 w \sqrt{E\rho}     }}  .
\end{align}
For an NV near the surface of the beam, $r_0 \sim t/2$ and we
obtain Eq.~(4) of the main text.

\subsection{Effective squeezing Hamiltonian from spin-phonon coupling}

In this section we provide additional details on
deriving $H_{\rm eff}$ in Eq.~(4) from
the original $H_{\rm NV}$ in Eq.~(1). 
Assuming that the magnetic field
is aligned along the NV axis,
$\vec B = B \hat z$, and defining $E_\pm = E_x \pm i E_y$
and  $S_\pm = S_x \pm i S_y$,
we can rewrite $H_{\rm NV}$ as ($\hbar = 1$)
\begin{align}
	H_{\rm NV} = ( D_0 + d_\parallel E_z ) S_z^2 + g_s \mu_B B S_z
	 -\frac{\dperp}{2} \left( E_+ S_+^2 + E_- S_-^2 \right),
	\label{eq:H1}
\end{align}
where $\vec E$ is the effective electric field
due to strain.
We quantize the perpendicular strain field, 
$E_+ = E_0 a$ and $E_- = E_0 a^\dagger$,
where $E_0$ is the strain due to the zero point motion
of the resonant mode.
Next we focus on  
the two-level subspace $\{ \ket{ 1} , \ket{-1}\}$ only, 
and assume that transitions to state $\ket{0}$ are not
allowed due to the large zero field splitting
$D_0$.  
For the $i$th spin we write Pauli operators
$\sigma^\pm_i = \ket{\pm 1}_i\bra{\mp 1}$
and 
$\sigma^z_i =   \ket{ 1}_i\bra{ 1} - \ket{- 1}_i\bra{- 1} $,
and within this two-level subspace the interaction for
a single NV is
\begin{equation}
\label{eq:Hnv2}
	H_i = \frac{\Delta_B}{2} \sigma^z_i + 
	g \left( \sigma^+_i a + a^\dagger \sigma^-_i \right) + \omega_m a^\dagger a,
\end{equation}
where $g = -d_\perp E_0$,
$\Delta_B = 2 g_s \mu_B B$ is the energy
between $\ket{\pm 1}$, and
we included the mechanical oscillator of frequency $\omega_m$.
Summing \eq{eq:Hnv2} for many NVs coupled to the same
mode with uniform coupling strength we obtain
\begin{equation}
\label{eq:HstartSI}
	H= \Delta_B J_z + 
	g \left( a^\dagger J_- + a J_+ \right) + \omega_m a^\dagger a,
\end{equation}
which is Eq.~(2) of the main text.
To obtain Eq.~(4), we first rewrite $H$
in the rotating frame at the mechanical frequency
$\omega_m$, 
\begin{equation}
	H= \Delta J_z + 
	g \left( a^\dagger J_- + a J_+ \right),
\end{equation}
where $\Delta = \Delta_B - \omega_m$.
Next we apply
the  transformation
$e^R H e^{-R}$, with
$R = \frac{g}{\Delta}  \left( a^\dagger J_- - a J_+ \right)$,
and to order $(g/\Delta)^2$ we obtain
\begin{equation}
	H_{\rm eff} \simeq \Delta J_z
	+ \frac{g^2}{\Delta} \left(
		J_+ J_- + 2 a^\dagger a J_z
	\right).
\end{equation}
Transforming back to the  nonrotating frame yields Eq.~(4)
of the main text.

\subsection{Individual spin dephasing and phonon-induced relaxation}
\label{sec:decoherence}

\subsubsection{Individual spin dephasing from intrinsic $T_2$}

Each NV spin experiences intrinsic decoherence
in the absence of the mechanical mode.
Individual  relaxation ($T_1$) processes are due
to lattice phonons; at low temperature $T_1$
can be $\sim 100$ s and we ignore it \cite{Jarmola2012supp}.
However, we include intrinsic single spin dephasing,
which arises 
from magnetic noise of $^{13}$C nuclear spins
in the diamond lattice.
In practice, single spin dephasing may be
nonexponential 
\cite{Taylor2008supp,deSousa2009supp},
but for simplicity
we  approximate the effect of 
single spin dephasing 
by an effective Markovian master equation 
with  dephasing rate $T_2^{-1}$,
\begin{equation}
\label{eq:T2}
	\dot \rho = \frac{1}{2T_2}
	\sum_i
	\left[ \sz_i \rho \sz_i - \rho \right].
\end{equation}

\subsubsection{Collective phonon-induced spin relaxation}

The transformation $R$ used to obtain the effective
squeezing Hamiltonian $H_{\rm eff}$
also introduces a relaxation channel for the collective
spin by admixing  phonon and spin degrees of freedom.
Mechanical dissipation is described by the
master equation for the system density matrix $\rho$,
\begin{equation}
\label{eq:dissOsc}
\begin{split}
	\dot \rho = &\gamma (\nth + 1) 
	\left[  a \rho a^\dagger - \shalf \left( a^\dagger a \rho + \rho a^\dagger a  \right) \right]\\
	&+ \gamma \nth
	\left[  a^\dagger \rho a - \shalf \left( a a^\dagger \rho + \rho a a^\dagger  \right) \right].
\end{split}
\end{equation}
Transforming $a$
and $a^\dagger$ using the transformation $R$ in the main text, 
we obtain
effective spin relaxation terms in the master equation,
\begin{equation}
\label{eq:dissSpin}
\begin{split}
	\dot \rho = &\Gamma_\gamma (\nth + 1) 
	\left[  J_- \rho J_+ - \shalf \left( J_+ J_- \rho + \rho J_+ J_-  \right) \right]
	\\&+ \Gamma_\gamma \nth
	\left[  J_+ \rho J_- - \shalf \left( J_- J_+ \rho + \rho J_- J_+  \right) \right]
\end{split}	
\end{equation}
where $\Gamma_\gamma =  \left(\frac{g}{\Delta}\right)^2 \gamma$.
From Eqs.~(\ref{eq:T2}) and (\ref{eq:dissSpin}) we calculate
the equations for spin  averages and variances accounting
for individual dephasing and collective relaxation using
$\partial_t \avg{A} = \tr{\left\{ A \dot \rho \right\}}$.

\subsubsection{Squeezing estimate
from linearized equations for spin averages and variances}
\label{sec:estimate}

Here we sketch the derivation of the estimated
optimal squeezing given in Eq.~(8) in the main text.
In order to  treat the squeezing Hamiltonian,
collective relaxation and  spin dephasing
on equal footing, we linearize the equations for the spin averages
and variances.
This corresponds to expanding in the small
error from decoherence at short times and 
ignoring the curvature of the Bloch sphere
for sufficiently short times, when the spin uncertainty distribution
remains on a locally flat region of the Bloch sphere.
To linearize the equations we assume that all 
(connected) correlations of order
higher than two vanish.
The linearized equations are  valid for short times,
so we also make use of the  initial conditions in the spin
coherent state at $t=0$, which are
$\avg{J_y} = \avg{J_z} = \avg{C_{yz}} = 0$
and $\avg{J_y^2} = \avg{J_z^2} = J/2$.
Here we define the covariance operator
$C_{yz} = (J_y J_z + J_z J_y) / 2$,
while in the main text we refer only to
its average,
$V_{yz} = \avg{C_{yz}}$.
Within these approximations, and using
Eqs.~(\ref{eq:T2}) and(\ref{eq:dissSpin}),  
the linearized
equations for the spin averages required
to calculate the squeezing parameter $\xi^2$ are
\begin{widetext}
\begin{align}
	\partial_t \avg{J_x}
	&= 
	-  \Gamma_2 \avg{J_x}
	\label{eq:JxFinal}
	\\
	\partial_t \avg{J_y}
	&= \lambda J \avg{J_z}  
	-\Gamma_2 \avg{J_y}
	- \Gamma_\gamma  ( \nth + \shalf) \avg{J_y} + 
			\Gamma_\gamma  \avg{C_{yz} } 
	\label{eq:JyFinal}
	\\
	\partial_t \avg{J_z} 
	&= -2 \Gamma_\gamma (\nth + \shalf) \avg{J_z} - 
	\Gamma_\gamma \left[ J (J+1) - \avg{J_z^2}\right]
	\label{eq:JzFinal}
	\\
	\partial_t \avg{J_y^2}
	&= 
	2 J\lambda \avg{C_{yz}}
	- 2 \Gamma_2 \left( \avg{J_y^2} - \frac{J}{2} \right)
	- 2\Gamma_\gamma  ( \nth + \shalf) \avg{J_y^2 - J_z^2} + 
			 \Gamma_\gamma J \avg{J_z} + \frac{\Gamma_\gamma}{2} \avg{J_z}
	\label{eq:Jy2Final}
	\\
	\partial_t \avg{J_z^2} 
	&= -2\Gamma_\gamma  (\nth + \shalf) \left[ 3 \avg{J_z^2} - J(J+1)\right]
	+ \Gamma_\gamma \avg{J_z} \left[ 1 - 2J(J+\shalf)\right]
	\label{eq:Jz2Final}
	\\
	\partial_t \avg{C_{yz}}
	&= 
	\lambda J \avg{J_z^2}
	- \Gamma_2 \avg{C_{yz}}
	- 5 \Gamma_\gamma  ( \nth + \shalf) \avg{C_{yz}} 
			- \Gamma_\gamma \left(J^2 - \tfrac{1}{4}\right) \avg{J_y}
	\label{eq:CyzFinal}
\end{align}
\end{widetext}
This linear set of equations can be directly solved.
The full analytic solutions are lengthy so we simply plot the
numerical solution for the squeezing parameter 
(blue dash-dotted in \fig{fig:numerics}), which
agrees with exact numerics at short times.

We use these linearized equations to
estimate the scaling
of the optimal squeezing parameter 
(see Eqs.~(7) and (8)  in the main text).
First, we solve
Eqs.~(\ref{eq:JxFinal}-\ref{eq:CyzFinal})
to second order in $t$.
Second, we calculate $\xi^2$
(see Eq.~(5) in the main text)
from the resulting spin averages, and
simplify the result in the limit of interest, $J \gg 1, \nth \gg 1$,
and  assuming 
$J \lambda t \gg 1$ as required for significant squeezing.
Third, we assume that all sources of decoherence
are small, and expand in the errors
$\Gamma_\gamma \nth t \ll 1$ and $ \Gamma_2  t \ll 1$.
Within these approximations we obtain 
\begin{equation}
	\xi^2 \simeq \frac{1 + 4 J \Gamma_\gamma \nth t}{\left(J\lambda t\right)^2} + 
		\left( 5 \Gamma_\gamma \nth + T_2^{-1} \right)  t,
\end{equation}
where the  first term $1 / (J \lambda t )^2$ is the result from linearized unitary squeezing,
and the remaining  terms are  the lowest order corrections
in both sources of decoherence.
We  further approximate $T_2^{-1} \gg \Gamma_\gamma \nth$,
valid for sufficiently large detuning $\Delta$, 
and $J\Gamma_\gamma \nth t \gg 1$,
valid 
self-consistently at the optimal squeezing time
and  in the relevant limit $J \gg \eta$.
Within these approximations we  obtain 
Eq.~(7) in the main text.
Finally, we optimize $\xi^2$ with respect to $t$,
obtaining 
Eq.~(8) and
$t_{\rm opt}$ given in the main text.

\subsubsection{Combining individual dephasing
and collective relaxation: numerics}

\begin{figure}[htb]
\centering
	\includegraphics[width=0.4\textwidth]{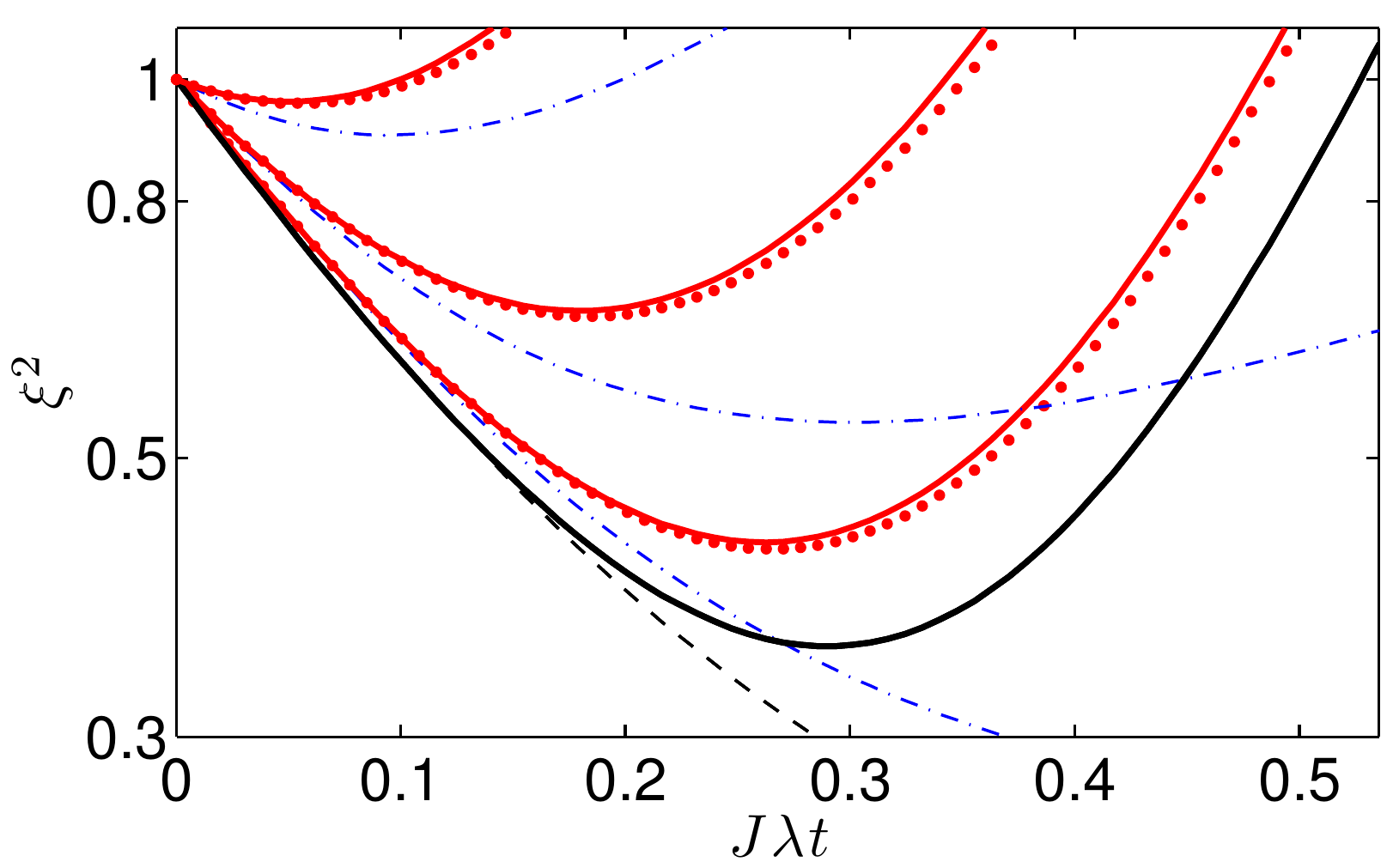}	
\caption{Spin squeezing parameter 
versus scaled precession time with $N = 8$ spins.
Solid red lines show squeezing calculated
using the approximation discussed in the text, 
treating single spin dephasing and collective relaxation
independently,
with
$\nth = 0, 10, 100$ (bottom to top).
Red dots show the exact numerics.
The detuning $\Delta$ is optimized for each
value of $\nth$.
Blue dash-dotted lines show squeezing
from linearized equations.
Solid black line shows unitary squeezing,
dashed black line shows unitary squeezing 
from linearized equations.
Parameters are
$\omega_m/2\pi = 1$ GHz,
$g / 2\pi = 1$ kHz,
$Q = 10^6$,
$T_2 = 100$ ms.}
\label{fig:numerics}
\end{figure}

As discussed in the main text, in the absence of a mechanical
drive we can neglect phonon number fluctuations for a sufficiently
large number of $\pi$ pulses $M$.
In this case
the remaining sources of decoherence
are intrinsic single spin dephasing
and collective relaxation induced by mechanical dissipation.
These sources of decoherence are simple to treat separately but
difficult to treat simultaneously for a large number of spins. 
To  calculate the solid blue curves in Fig. 2 of the main text,
we treat the combination of both  sources of decoherence as
approximately independent,
valid provided both are small enough to still allow
spin squeezing.
To calculate the spin averages needed
for the squeezing parameter, we first account for
collective relaxation using the Dicke state basis
in which total $J$ is conserved.
We then account for individual dephasing by multiplying the
resulting 
averages by dephasing factors such as 
$\avg{J_x(t)} = e^{-t/T_2} \avg{J_x(t)}_{\rm D}$
where $\avg{J_x(t)}_{\rm D}$ is the   result of 
the Dicke state
calculation.
Each step  would be numerically exact
in the absence of the other source of decoherence;
thus we expect that this procedure provides a good approximation
if all errors are small.
To verify the accuracy of the approach, we compare the result
with exact numerics calculated by numerically
integrating the full master equation
for small $N$ in \fig{fig:numerics}.

\subsection{Phonon number fluctutions}

In this section we consider  fluctuations of the phonon number,
$n = a^\dagger a$.
We start by rewriting the effective Hamiltonian 
[see Eq.~(4) in main text]
for the collective spin coupled to a  driven oscillator
in the  frame rotating at the mechanical drive frequency,
\begin{equation}
	H_{\rm eff} =  \Delta J_z
	+ \lambda  J_z a^\dagger a + \frac{\lambda}{2} J_z^2 
	-\delta a^\dagger a
	+ \Omega (a+a^\dagger)
\end{equation}
where $\Delta = \Delta_B - \omega_d$ is the detuning
of the magnetic transition frequency from the drive,
and $\delta = \omega_d - \omega_m$ is the drive detuning
from the mechanical frequency.
The amplitude of the drive is $\Omega$ and we have
made the rotating wave approximation. 
Our aim is
to find the effect of the oscillator on the spin
to second order in $\lambda$
(within the Gaussian approximation).
For this we require
the number fluctuation spectrum of  a 
damped, driven,
thermal oscillator in the absence of coupling to the spin.

\subsubsection{Number fluctuations of a driven thermal mode}
\label{app:Sn}

To calculate the effective dephasing from number fluctuations,
we  first need 
the power spectral density of phonon number fluctuations,
\begin{equation}
\label{eq:Sndef}
	S_n(\omega) = \int dt e^{i \omega t} \avg{\delta n(t) \delta n(0)},
\end{equation}	
where $n = a^\dagger a$, $\delta n = n - \avg{n}$
and the average is taken with respect to the oscillator in thermal
equilibrium with its environment.
In the absence of coupling, $\lambda = 0$,
the Langevin equation for the driven thermal mode
in the  frame of the classical drive frequency and
within the rotating wave approximation is
\begin{equation}
	\dot a(t) = \left( i\delta - \frac{\gamma}{2} \right) a(t)
	+ \Omega + \sqrt \gamma \xi(t)
\end{equation}
The solution is $a(t) = \alpha + d(t)$,
where $\alpha = \frac{ \Omega}{ - i \delta + \gamma / 2}$
is the coherent amplitude due to the drive, and
\begin{equation}
	d(t) = 
	\sqrt{\gamma} \int_{-\infty}^t dt' e^{i \delta - \gamma/2)(t-t')} \xi(t') 
\label{eq:d}
\end{equation}
describes thermal and quantum fluctuations.
The mean phonon number is the sum
of driven and thermal parts,
$\n = \ndr + \nth$, where
\begin{equation}
	\ndr = \abs{\alpha^2} = \frac{\Omega^2}{\delta^2 + \gamma^2/4} \\
\end{equation}
and the thermal occupation is $\nth = \avg{d^\dagger d} = 1 / (e^{\omega_m/T} - 1)$.
Using \eq{eq:d} we find the two-time correlations,
\begin{align}
	\avg{d^\dagger(t)d(0)} &= \nth e^{-i\delta t} e^{-\gamma \abs{t} / 2},
	\\
	\avg{d(t)d^\dagger(0)} &= (\nth+1) e^{i\delta t} e^{-\gamma \abs{t} / 2},
\end{align}
and from these 
we can calculate the full spectrum of driven thermal number fluctuations.
The  correlation using Wick's theorem 
and $a(t) = \alpha + d(t)$ is
\begin{align}
	\redavg{\delta n(t) \delta n(0)} =& \avg{n(t) n(0)} - \n^2
	\nonumber \\
	=& \ndr \left[ \avg{d^\dagger(t) d(0)}
	+ \avg{d(t)d^\dagger(0)} \right]\\
	&+ \avg{d^\dagger(t)d(0)}\avg{d(t)d^\dagger(0)}.
\end{align}
Using  \eq{eq:d} and taking the Fourier transform, we find
that 
the number fluctuation spectrum for a driven
thermal oscillator
is given by
\begin{equation}
\label{eq:Sn}
\begin{split}
	S_n(\omega) 
	= &\gamma  \ndr
	\left[
	\frac{\nth}{(\omega-\delta)^2 + \gamma^2/4}
	+ 	\frac{\nth + 1}{(\omega+\delta)^2 + \gamma^2/4}
	\right]\\
	&+ \frac{2\gamma \nth (\nth + 1)}{\omega^2 + \gamma^2}.
\end{split}
\end{equation}

\subsubsection{Effects of number fluctuations 
on the spin in the Gaussian approximation}

From
the spectrum of number fluctuations we can
calculate the effect of 
number fluctutions
on the spin dephasing and squeezing.
We write the full Hamiltonian as
$H_{\rm eff} = H_0 + H_{\rm osc} + \Hint$, where $H_{\rm osc}$
describes the driven damped oscillator, and
\begin{align}
	H_0 = \Delta  J_z + \frac{\lambda}{2} J_z^2
\end{align}
describes the spin including the constant 
effective squeezing term.
The coupling in the interaction picture 
and in the toggling frame  is
\begin{equation}
	\Hint(t)  =  \lambda f(t) \delta n(t) J_z,
\end{equation}
where $J_z$ is time-independent
as it commutes with the full Hamiltonian.
We have included the function $f(t)$ to describe
spin echo, which effectively inverts the sign
of the interaction with each $\pi$ pulse.

The equation of motion for the operator $J_+$ in the interaction picture is
\begin{equation}
	\dot J_+(t) = i \lambda f(t) \big[ J_z \delta n(t) , J_+(t) \big].
\end{equation}
We integrate this  formally, insert the solution, 
and take the average with respect to the oscillator to get
\begin{equation}
	\dot J_+(t) = - \lambda^2 f(t) \int_0^t f(t') 
	\avg{\big[ J_z \delta n(t) \comm{J_z \delta n(t') , J_+(t')}  \big]}_{\rm osc}
\end{equation}
where $\avg{\cdot}_{\rm osc}$ denotes averaging  
over the oscillator degrees of freedom.
Note that we neglected additional noise terms;
these play no role as we
will only be interested in taking the average at the end.
Next, we neglect the time dependence of $J_+(t')$ under
the integral, as it is higher order in $\lambda$,
$J_+(t) = e^{i H_0 t} J_+ e^{-i H_0 t} = e^{i \lambda (J_z - 1/2) t} J_+$.
Expanding the commutators we obtain
\begin{widetext}
\begin{align}
	\dot J_+ &= - \lambda^2 f(t) \int_0^t f(t') 
	\left[
	J_z J_+ \avg{\delta n(t) \delta n(t')}
	- J_+ J_z \avg{\delta n(t') \delta n(t)}
	\right] \nonumber \\
	&= 
	- \lambda^2 f(t) \int_0^t f(t') 
	\int \frac{d\omega}{2\pi} e^{-i \omega (t-t')}
	\left[
	J_z J_+ S_n(\omega)
	- J_+ J_z S_n(-\omega)
	\right],
	\label{eq:JplusDot}
\end{align}
using \eq{eq:Sndef}.
Defining the symmetric and antisymmetric
parts of the number fluctuation spectrum,
\begin{equation}
	\bar S_n(\omega) 
	= \half \left[ S_n(\omega) + S_n(-\omega) \right]
	,\quad
	A_n(\omega) 
	= S_n(\omega) - S_n(-\omega),
\end{equation}
we can rewrite \eq{eq:JplusDot} as
\begin{equation}
	\dot J_+ = 
	- \lambda^2 f(t) \int_0^t f(t') 
	\int \frac{d\omega}{2\pi} e^{-i \omega (t-t')}
	\times\left[
	\bar S_n(\omega)
	+  A_n(\omega) \left( J_z - \frac{1}{2} \right) 
	\right] J_+.
\label{eq:JplusDotFinal}
\end{equation}
Solving \eq{eq:JplusDotFinal} and finally taking the average with respect
to  spin degrees of freedom, we obtain
\begin{equation}
	\avg{J_+(t)} = e^{-\chi} \avg{e^{i \mu (J_z - \frac{1}{2})} J_+(0)}
\end{equation}
where $\avg{\cdot}$ is the average over all degrees of freedom, and
\begin{align}
	\chi
	&= \lambda^2 
	\int \frac{d\omega}{2\pi}
	 \bar S_n(\omega)
	 \int dt_1 
	 \int dt_2
	 e^{-i\omega(t_1-t_2)}
	 \theta(t_1-t_2)
	 f(t_1)
	  f(t_2),
	  \\
	  \mu
	 &=
	   i \lambda^2 
	 \int \frac{d\omega}{2\pi}
	  A_n(\omega)
	 \int dt_1 
	 \int dt_2
	 e^{-i\omega(t_1-t_2)}
	 \theta(t_1-t_2)
	 f(t_1)
	  f(t_2).
\end{align}
\end{widetext}
Here all integration limits are  from $-\infty$ to $\infty$,
and the time integration limits are accounted for in 
$f(t') \propto \theta(t')\theta(t-t')$
and
the step function $\theta(t_1 - t_2)$.
Similarly, we obtain the other averages needed to
calculate the squeezing,
\begin{align}
	\avg{J_+^2(t)}
	&= e^{-4 \chi} \avg{e^{2i\mu(J_z-1)} J_+^2(0)},
	\\
	\avg{J_+(t)J_z(t)}
	&= e^{- \chi} \avg{e^{i \mu \left(J_z - \frac{1}{2}\right)} J_+(0)J_z(0)}.
\end{align}
By comparing with the spin evolution under
unitary one-axis twisting,
we see that  $\mu$ 
describes  spin squeezing with an effective
squeezing coefficient $\lambda_{\rm eff}  = \mu / t$.
The parameter $\chi$ describes collective dephasing.

To evaluate $\chi$ and $\mu$ for a given pulse sequence,
we next  define $f(\omega) = \int dt e^{i\omega t} f(t)$
to rewrite the double time integral as
\begin{align}
	I_t =&
	\int dt_1 
	 \int dt_2
	 e^{-i\omega(t_1-t_2)}
	 \theta(t_1-t_2)
	 f(t_1)
	  f(t_2)\\
	 &=
	  \int \frac{d\omega_1}{2\pi}
	  \abs{ f(\omega_1)}^2
	\left[
	\pi \delta(\omega+\omega_1) - \frac{i}{\omega+\omega_1}
	\right].
\end{align}	
We define the filter function for pulse sequence
with time $\tau$ between $\pi$ pulses,
\begin{equation}
	F(\omega\tau) = \frac{\omega^2}{2} \abs{f(\omega)}^2.
\end{equation}
The dephasing term is
\begin{align}
	\chi
	&= 2\lambda^2 \int \frac{d\omega}{2\pi}
	\bar S_n(\omega)  \int \frac{d\omega_1}{2\pi}
	  \frac{F(\omega_1 \tau)}{\omega_1^2}
	\left[
	\pi \delta(\omega+\omega_1) - \frac{i}{\omega+\omega_1}
	\right]
\end{align}
Since $F(\omega\tau)$ and $\bar S_n(\omega)$ are
both {even} in $\omega$,
the imaginary part of the
integrand is odd and integrates to
zero.
As  a result $\chi$ is real and we obtain
\begin{equation}
\label{eq:chiSI}
	\chi =
	\lambda^2 \int \frac{d\omega}{2\pi}
	\frac{F(\omega \tau)}{\omega^2}
	\bar S_n(\omega).
\end{equation}

The coherent term is
\begin{align}
	\mu
	&=  i \lambda^2 \int \frac{d\omega}{2\pi}
	 A_n(\omega)  \int \frac{d\omega_1}{2\pi}
	  \frac{F(\omega_1 \tau)}{\omega_1^2}
	\left[
	\pi \delta(\omega+\omega_1) - \frac{i}{\omega+\omega_1}
	\right].
\end{align}
Since $A_n(\omega)$ is odd, 
in this case the imaginary part (involving the $\delta$-function)
is zero.
The real part is
\begin{equation}
\label{eq:muSI}
	\mu =
	\lambda^2 \int \frac{d\omega}{2\pi}
	\frac{K(\omega \tau)}{\omega^2}
	A_n(\omega),
\end{equation}
where we defined 
\begin{equation}
	K(\omega\tau) = 2\omega^2 \int \frac{d\omega_1}{2\pi}
	\frac{F(\omega_1\tau)}{\omega_1^2 (\omega+\omega_1)}.
\end{equation}
$K(\omega\tau)$ and $F(\omega\tau)$ 
satisfy a Kramers-Kronig relation
(with a factor of $\omega^2$ from the definitions).

\subsubsection{Dephasing from purely thermal oscillator}

From \eq{eq:Sn}, the number fluctuation
spectrum of a purely thermal oscillator in the
frame of the mechanical drive is
\begin{equation}
\label{eq:SnThermal}
	S_n^{\rm th}(\omega) 
	= 
	 \frac{2\gamma \nth (\nth + 1)}{\omega^2 + \gamma^2}.
\end{equation}
The spectrum is symmetric in frequency and thus
$\mu = 0$.
We obtain the dephasing from  thermal flucuations
by inserting \eq{eq:SnThermal} in \eq{eq:chiSI}.
For a  sequence with an even number   of pulses $M$,
the filter function is 
\begin{equation}
	F_{M}(\omega\tau) = 2 \sin^2\left(\frac{M \omega\tau}{2}\right)
		\left[ 1 - \sec\left(\frac{\omega\tau}{2}\right)\right]^2,
\end{equation}
where $\tau$ is the time between evenly spaced pulses and the total
sequence time is $t = N\tau$.
In the relevant limit $\gamma \tau \ll 1$
we obtain $\chi_{\rm th}$
given in the main text.

\subsubsection{Dephasing and squeezing from driven thermal oscillator}

Adding a mechanical drive, the total dephasing from number fluctuations
becomes $\chi = \chi_{\rm th} + \chi_{\rm dr}$, where $\chi_{\rm dr}$ is
obtained from the driven part of the  
number fluctuations,
\begin{equation}
\label{eq:SnDriven}
	 S_n^{\rm dr}(\omega) 
	= \gamma  \ndr
	\left[
	\frac{\nth}{(\omega-\delta)^2 + \gamma^2/4}
	+ 	\frac{\nth + 1}{(\omega+\delta)^2 + \gamma^2/4}
	\right].
\end{equation}
The dephasing involves the symmetrized part,
 \begin{equation}
\label{eq:SnDrivenSymmetric}
\begin{split}
	\bar  S_n^{\rm dr}(\omega) 
	=& \gamma  \ndr \left( \nth + \frac{1}{2} \right)\\
	&\times \left[
	\frac{1}{(\omega-\delta)^2 + \gamma^2/4}
	+ 	\frac{1}{(\omega+\delta)^2 + \gamma^2/4}
	\right].
\end{split}
\end{equation}
Using \eq{eq:chiSI} this yields the dephasing from a
driven for an $M$ pulse  sequence.
In the relevant limit $\gamma\tau \ll 1$ and
choosing the timing $\tau = 2\pi / \delta$, we obtain
$\chi_{\rm dr}$
given in the main text.

For a driven oscillator the power spectral density is {\it not} symmetric,
and the asymmetric part can lead to additional squeezing.
The asymmetric part of $S_n(\omega)$ is 
\begin{equation}
	A_n(\omega) = \frac{\gamma \ndr}{2}
	\left[  
	\frac{1}{(\omega+\delta)^2 + \gamma^2/4}
	-	\frac{ 1}{(\omega-\delta)^2 + \gamma^2/4}
	\right].
\end{equation}	
Using \eq{eq:muSI},
and choosing the pulse timing to coincide with a 
coherence ``revival", $\tau = 2\pi / \delta$, and assuming
the mechanical $Q \gg 1$, we obtain
$\mu_{\rm dr}$
given in the main text.

\subsubsection{Optimized squeezing with drive}

With a strong mechanical drive, the 
approximate optimal squeezing is obtained similarly as 
in Sec.~\ref{sec:estimate} above.
In the driven case we assume that the detuning is large, so that
$\Gamma_\gamma \rightarrow 0$, and
the drive is strong so that $\ndr \gg \nth$ and $\chi_{\rm dr} \gg \chi_{\rm th}$.
Again expanding $\xi^2$ 
in the limit $J \gg 1$, $\nth \gg 1$, and
small errors $t/T_2, \chi_{\rm dr} \ll 1$,
we obtain
\begin{equation}
	\xi^2(t) \simeq  \frac{ \gamma\nth }{J g^2 t} 
		+ \frac{t}{T_2}
\end{equation}
where we chose
$\ndr \sim (\delta/g)^2$,
 the maximum allowed driving strength in our perturbative treatment
of the coupling.
Optimizing with respect to $t$ we recover Eq.~(8) in the main text.

\end{document}